\documentclass[aps,prx,superscriptaddress,twocolumn,nofootinbib,10pt]{revtex4-2}

\usepackage[utf8]{inputenc}
\usepackage[titletoc,toc,title,page]{appendix}
\usepackage{csquotes}
\usepackage[english]{babel}
\usepackage{microtype}
\usepackage{bm}
\usepackage{dsfont}
\usepackage{amsmath,amssymb,amsthm,thmtools}
\usepackage{mathtools}
\usepackage{cases}
\usepackage{calc}
\usepackage{mathrsfs}
\usepackage[normalem]{ulem}
\usepackage{parskip}
\usepackage[usenames,dvipsnames,table]{xcolor}
\usepackage{nameref}
\usepackage[colorlinks=true]{hyperref}
\usepackage[nameinlink]{cleveref}
\usepackage{physics}
\usepackage{float}
\usepackage{graphicx}
\usepackage{easyReview}
\usepackage{tikz}
\usetikzlibrary{calc,shapes.geometric}
\usepackage{placeins}
\usepackage{multirow,tabularx,booktabs}
\usepackage[most]{tcolorbox}
\usepackage[printonlyused,withpage,nohyperlinks,smaller]{acronym}
\usepackage{enumitem}

\crefname{appsec}{Appendix}{Appendices}
\crefname{box}{Box}{Box}
\hypersetup{
  colorlinks = true,
  urlcolor   = green!80!black,
  linkcolor  = blue,
  citecolor  = red!80!black
}

\graphicspath{{./figures/}}

\newtcbtheorem{tbox}{Box}{enhanced, float*=t, width=\textwidth, label type=box}{box}

\newcommand{\bs}[1]{\boldsymbol{#1}}

\begin{document}

\title{Quantum Fisher information and imperfect detection in a monitored fermion chain}
\author{Giovanni Di Fresco}
\affiliation{\resizebox{0.95\textwidth}{!}{SISSA-International School for Advanced Studies and INFN, Sezione di Trieste, via Bonomea 265, 34136 Trieste, Italy}}
\author{Davide Valenti}
\affiliation{%
	\begin{tabular}[t]{c}
		\mbox{Università degli Studi di Palermo, Dipartimento di Fisica e Chimica--Emilio Segré,}
		\mbox{Group of Interdisciplinary Theoretical Physics,}\\
		\mbox{Università degli Studi di Palermo, Viale delle Scienze, Ed.~18, I-90128 Palermo, Italy}
	\end{tabular}%
}
\author{Angelo Carollo}
\affiliation{%
	\shortstack[c]{%
		Università degli Studi di Palermo, Dipartimento di Fisica e Chimica--Emilio Segré,
		via Archirafi 36, I-90123 Palermo, Italy%
	}%
}

\begin{abstract}
We study the metrological properties of a continuously monitored Kitaev chain in the presence of imperfect detection. The system is conditioned on a no-click record, while each emitted fermion is registered only with probability $0\leq q\leq 1$. Because the conditional dynamics remains Gaussian, the steady state is fully characterized by the fermionic correlation matrix. This allows a direct evaluation of the quantum Fisher information and of the mean Uhlmann curvature. For perfect detection, the monitored steady state retains a singular critical structure and the quantum Fisher information with respect to the chemical potential becomes super-extensive. For any $q<1$, imperfect detection introduces a finite smoothing length that rounds the singularity and restores extensive scaling. The detector efficiency behaves instead as a compatible mixed-state estimation parameter, as signaled by the vanishing mean Uhlmann curvature. These results show that incomplete trajectory information destroys the metrological enhancement associated with monitored criticality through a mechanism that differs from ordinary thermal smearing.
\end{abstract}

\maketitle

\section{Introduction}
Continuous monitoring can qualitatively reshape the many-body dynamics of quantum systems. In monitored free-fermion and spin chains, the competition between coherent evolution and measurement backaction gives rise to measurement-induced transitions in entanglement, correlations, and steady-state structure \cite{Alberton2020Entanglement,Fuji2020Measurement,Zerba2023NoClick, li2018,skinner2019,gullans2020,koh2023,cao2019,choi2020,block2022,minato2022,zabalo2020,li2019,sharma2022,bao2020,turkeshi2021,piccitto2022,barratt2022,coppola2022,fux2023,passarelli2024}. In the no-click limit, these transitions admit a sharp interpretation in terms of non-Hermitian conditional dynamics~\cite{Biella2021} and can produce singular behavior in the correlation matrix together with critical scaling of information-theoretic observables \cite{Zerba2023NoClick,Paviglianiti2024QFI,DiFresco2023MIPTMetrology}.
A central issue is the robustness of this picture under imperfect detection. In realistic protocols, only a fraction of the emitted excitations is actually registered, so the observer conditions the dynamics on incomplete trajectory information. In monitored fermionic chains, this loss of information suppresses the critical long-range structure of the ideal no-click dynamics by generating a finite correlation length, thereby blurring the signatures of the measurement-induced transition \cite{Paviglianiti2024Breakdown,Ladewig2022Monitored}. The effect is best viewed as partial averaging over quantum trajectories rather than as ordinary unconditional dissipation~\cite{Paviglianiti2024Breakdown,Coppola2023Conditional}.
\newline
Here we study the metrological consequences of this mechanism in a continuously monitored Kitaev chain with fermionic losses and finite detector efficiency $q$. Because the conditional dynamics remains Gaussian, the steady state is fully characterized by the fermionic correlation matrix \cite{Coppola2023Conditional,Carollo2018SLD}. This makes it possible to analyse both the quantum Fisher information with respect to the chemical potential and the mean Uhlmann curvature associated with joint estimation of $\mu$ and $q$. The main question is whether the critical metrological enhancement known to accompany monitored criticality survives once imperfect detection rounds the singularity.
\emph{We show that it does not}. At perfect detection, the monitored steady state retains a singular critical structure and the quantum Fisher information is super-extensive. For any $q<1$, imperfect detection introduces a finite smoothing length that restores extensive scaling. At the same time, the detector efficiency behaves as a compatible estimation parameter, as signaled by the vanishing mean Uhlmann curvature \cite{Carollo2017Uhlmann,DiFresco2022Multiparameter,Ragy2016Compatibility}. The result identifies incomplete monitoring as a metrologically relevant mechanism that preserves mixed-state compatibility, while destroying the Fisher enhancement tied to monitored criticality.
\begin{figure}
    \centering
    \includegraphics[width=1\linewidth]{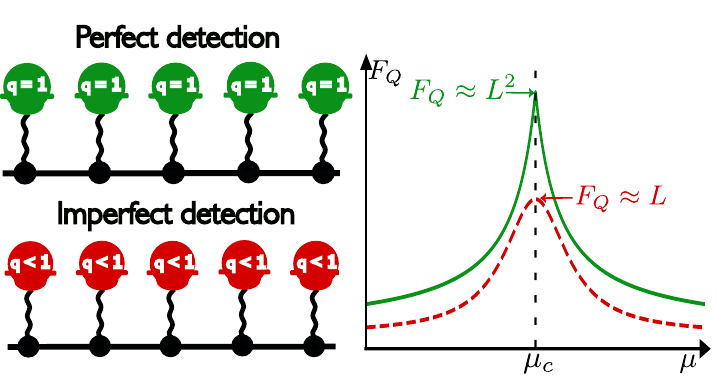}
    \caption{Schematic illustration of the effect of imperfect detection on the Fisher-information peak near criticality. For perfect detection, $q=1$, the Fisher information develops a sharp singular contribution at the critical point $\mu_c$, solid green line. For imperfect detection, $q<1$, this singularity is rounded into a smooth crossover, red dashed line, reflecting the smoothing of the critical mode and the restoration of extensive scaling.} 
    \label{fig:sk}
\end{figure}
\section{Model and conditional Gaussian dynamics}
We consider a Kitaev chain of length $L$ subject to local fermionic losses and conditioned on a no-click measurement record with finite detector efficiency $0\leq q\leq 1$. This is the imperfect-detection extension of the monitored Kitaev-chain setting analysed in Ref.~\cite{Paviglianiti2024Breakdown}. The conditional state is generally mixed for $q<1$, because the observer postselects on an incomplete record rather than on a pure trajectory. For quadratic Hamiltonians and linear jump operators, however, the dynamics remains Gaussian and can be written entirely in terms of the fermionic correlation matrix~\cite{Coppola2023Conditional,Genoni2016Conditional,Barthel2021Solving}.
\newline
The conditional evolution is described by
\begin{subequations}\label{eq:liouvillian_system}
\begin{equation}\label{eq:master_equation}
\partial_t \hat{\rho}_t = \mathcal{L}_t \hat{\rho}_t,
\end{equation}
\begin{align}\label{eq:liouvillian}
\mathcal{L}_t\,\bullet &= -i\comm{\hat{H}}{\bullet} + \\
&\quad + \sum_{j=1}^{L}\Bigg[(1-q)\hat{L}_j\bullet\hat{L}_j^{\dagger}
-\frac{1}{2}\acomm{\hat{L}_j^{\dagger}\hat{L}_j}{\bullet}  + q\expval{\hat{L}_j^{\dagger}\hat{L}_j}_t\,\bullet\Bigg].\notag
\end{align}
\end{subequations}
where $\hat{L}_j=\sqrt{\gamma}\,\hat{c}_j$ describes on-site fermionic loss and $\gamma$ is the monitoring strength \cite{Paviglianiti2024Breakdown, Minganti2020}. The lattice Hamiltonian is the Kitaev chain
\begin{equation}\label{eq:kitaev_hamiltonian}
\hat{H}=-\sum_{j=1}^{L}\left(\hat{c}_j^{\dagger}\hat{c}_{j+1}+\hat{c}_j^{\dagger}\hat{c}_{j+1}^{\dagger}+\mathrm{h.c.}\right)+2\mu\sum_{j=1}^{L}\hat{c}_j^{\dagger}\hat{c}_j,
\end{equation}
with periodic boundary conditions. The nonlinear term in Eq.~\eqref{eq:liouvillian} ensures trace preservation of the conditional state and is the characteristic signature of no-click conditioning with imperfect detection \cite{Paviglianiti2024Breakdown,Coppola2023Conditional}.
\newline
We introduce Majorana operators
\begin{equation}
\hat{w}_{j,1}=\frac{\hat{c}_j+\hat{c}_j^{\dagger}}{\sqrt{2}},
\qquad
\hat{w}_{j,2}=-i\frac{\hat{c}_j-\hat{c}_j^{\dagger}}{\sqrt{2}},
\end{equation}
and define the fermionic correlation matrix as
\begin{equation}\label{eq:gamma_def}
\Gamma_{(i,\alpha),(j,\beta)}=\frac{i}{2}\Tr\left(\rho\,[\hat{w}_{i,\alpha},\hat{w}_{j,\beta}]\right).
\end{equation}
Because the Hamiltonian is quadratic and the jump operators are linear, Gaussianity is preserved throughout the evolution \cite{Coppola2023Conditional,Genoni2016Conditional}. Writing
\begin{equation}
\hat{H}=\sum_{m,n=1}^{L}\sum_{\alpha,\beta=1}^{2}\mathbb{H}_{(m,\alpha),(n,\beta)}\hat{w}_{m,\alpha}\hat{w}_{n,\beta},
\end{equation}
with $\mathbb{H}=-\mathbb{H}^{T}$, and
\begin{equation}
\hat{L}_j=\sum_{m=1}^{L}\sum_{\alpha=1}^{2}\ell^{(j)}_{m,\alpha}\hat{w}_{m,\alpha},
\end{equation}
we define the bath matrix
\begin{equation}\label{eq:bath_matrix}
M_{(m,\alpha),(n,\beta)}=\sum_{j=1}^{L}\ell^{(j)}_{m,\alpha}\left(\ell^{(j)}_{n,\beta}\right)^*.
\end{equation}
The correlation matrix then obeys a Riccati-type equation \cite{Coppola2023Conditional}
\begin{equation}\label{eq:time_ARE}
\partial_t\Gamma_t=X\Gamma_t+\Gamma_tX^T+Y+\Gamma_tZ\Gamma_t,
\end{equation}
with
\begin{subequations}
\begin{align}
X&=-2i\mathbb{H}-(1-q)\Re M,\\
Y&=\left(1-\frac{q}{2}\right)\Im M,\\
Z&=2q\Im M.
\end{align}
\end{subequations}

For a translationally invariant chain it is convenient to work in momentum space. Writing $A\in\{\Gamma,X,Y,Z\}$ and defining $\tilde{A}(k)$ as the Fourier transform of the relative-coordinate kernel, Eq.~\eqref{eq:time_ARE} decouples into independent equations for each momentum mode. In the steady state one obtains the $2\times 2$ algebraic Riccati equation \cite{Paviglianiti2024Breakdown}
\begin{equation}\label{eq:ARE}
\tilde{X}(k)\tilde{\Gamma}(k)+\tilde{\Gamma}(k)\tilde{X}^T(-k)+\tilde{Y}(k)+\tilde{\Gamma}(k)\tilde{Z}(k)\tilde{\Gamma}(k)=0.
\end{equation}

The exact steady-state solution follows Ref.~\cite{Paviglianiti2024Breakdown}. Imposing the physical constraints
\begin{equation}
\tilde{\Gamma}_{11}(k)=-\tilde{\Gamma}_{22}(k),
\qquad
\tilde{\Gamma}(k)=-\tilde{\Gamma}^{\dagger}(k),
\end{equation}
one can parameterize the solution as
\begin{equation}\label{eq:Gamma_param}
\tilde{\Gamma}(k)=iA(k)
\begin{pmatrix}
1 & a(k)+ib(k)\\
a(k)-ib(k) & -1
\end{pmatrix}.
\end{equation}
The explicit closed-form expressions for the two algebraic branches are given in Appendix~\ref{app:closed_form_solution}. In the main text we only retain the physical content that matters for the metrological analysis: among the admissible branches, the steady state is selected by requiring the correct limits as $q\to 0$ and $q\to 1$ \cite{Paviglianiti2024Breakdown}. In what follows we denote this physical branch simply by $\tilde{\Gamma}(k)$.

\section{Quantum Fisher information and compatibility}
The relevant precision bounds are set by the quantum Cram\'er--Rao framework \cite{Paris2008Estimation}. For a family of states $\rho(\bs{\theta})$ depending on parameters $\bs{\theta}=(\theta_1,\theta_2,\ldots)$, the covariance matrix of any locally unbiased estimator satisfies
\begin{equation}
\mathrm{Cov}(\bs{\theta})\geq F_Q^{-1},
\end{equation}
where the QFI matrix is defined by
\begin{equation}\label{eq:QFI_def}
(F_Q)_{\alpha\beta}=\frac{1}{2}\Tr\left[\rho(\bs{\theta})\left\{\Lambda_{\alpha},\Lambda_{\beta}\right\}\right],
\end{equation}
and the symmetric logarithmic derivatives satisfy
\begin{equation}
2\partial_{\alpha}\rho(\bs{\theta})=\left\{\Lambda_{\alpha},\rho(\bs{\theta})\right\}.
\end{equation}
In many-body systems the QFI is usually extensive away from criticality and can become super-extensive near a continuous transition, reflecting the singular sensitivity of the state to parameter changes \cite{Zanardi2007Resource,Invernizzi2008Optimal,Frerot2018}. The same logic extends to monitored criticality, where the Fisher information displays nonanalytic behavior across measurement-induced transitions~\cite{DiFresco2023MIPTMetrology, Paviglianiti2024QFI}.
\newline
For fermionic Gaussian states the SLD and the QFI can be expressed directly in terms of the correlation matrix and its derivatives \cite{Carollo2018SLD}. In the basis that diagonalizes $\Gamma$, with eigenvalues $c_r$, one can write
\begin{equation}\label{eq:F_gau}
\frac{(F_Q)_{\alpha\beta}}{4}=\sum_{rs}^{\prime}\frac{(\partial_{\alpha}\Gamma)_{rs}(\partial_{\beta}\Gamma)_{sr}}{1-c_rc_s},
\end{equation}
where the prime indicates that terms with $c_rc_s=1$ are excluded.
\newline
To discuss joint estimation of $\mu$ and $q$, we also consider the mean Uhlmann curvature \cite{Carollo2017Uhlmann,Carollo2019Quantumness}
\begin{equation}\label{eq:MUC_def}
U_{\alpha\beta}=\frac{i}{4}\Tr\left[\rho\,[\Lambda_{\alpha},\Lambda_{\beta}]\right],
\end{equation}
which, in the Gaussian setting, can be expressed in terms of the same correlation-matrix data entering the quantum Fisher information as
\begin{equation}
U_{\alpha\beta}=4\sum_{ij}^{\prime}\frac{c_i-c_j}{1-c_ic_j}\,\partial_{\alpha}\Gamma_{ij}\partial_{\beta}\Gamma_{ji}.
\label{eq:MUC}
\end{equation}
Its vanishing is the standard signature that the discrepancy between the QFI Cram\'er--Rao bound and the ultimately attainable multiparameter bound disappears \cite{Ragy2016Compatibility,DiFresco2022Multiparameter}.

\section{Perfect-detection critical scaling}
\begin{figure}
    \centering
    \includegraphics[width=0.9\linewidth]{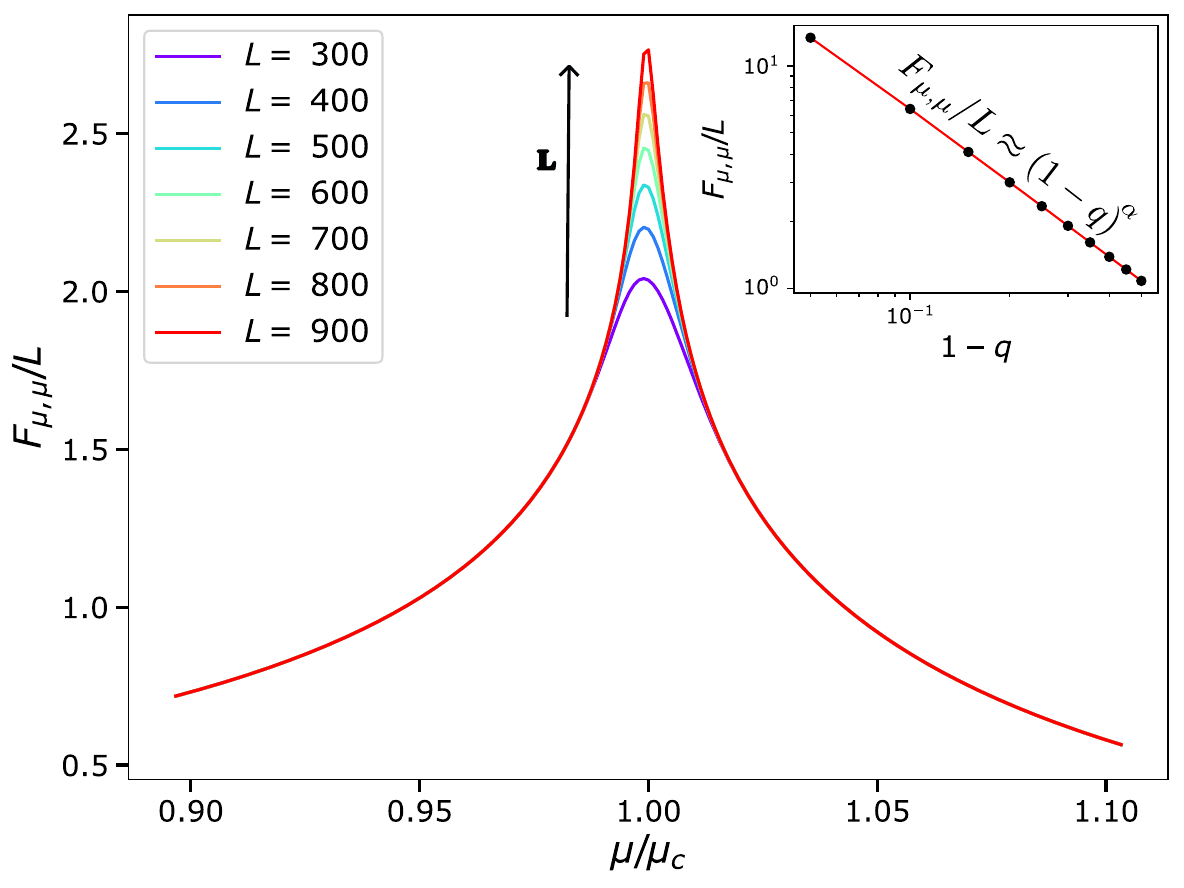}
    \caption{The main plot shows $F_{\mu\mu}$ for $q=1$ across the critical point, clearly displaying the superextensive peak at criticality. The inset shows the scaling behavior of the Fisher information away from perfect detection on a log-log scale, with a fitted exponent $\alpha = -1.095$. In these simulations, $\gamma = 1$.} 
    \label{fig:FQ_vs_N}
\end{figure}

We first discuss the ideal case $q=1$. Because Eq.~\eqref{eq:F_gau} is invariant under unitary transformations, the diagonal QFI component associated with the chemical potential can be resolved mode by mode in momentum space. For the present $2\times 2$ correlation matrix, the mode-resolved expression and the corresponding Bloch-vector decomposition are reported in Appendix~\ref{app:qfi_mode_decomposition}.
\newline
At perfect detection, the conditional steady state is pure and the correlation matrix develops a discontinuity at a critical momentum $k^*(\mu)$. The derivative with respect to $\mu$ is therefore dominated by the singular contribution of that mode, which yields a super-extensive piece in the QFI. The detailed derivation is given in Appendix~\ref{app:scaling_arguments}. Here we quote the resulting scaling law,
\begin{equation}\label{eq:L2_scaling}
F_{\mu\mu}=\frac{B L^2}{\pi^2\left(1-\mu^2\right)},
\end{equation}
where $B$ is a nonuniversal prefactor set by the jump amplitude and the size of the discontinuity. This scaling law is further confirmed by a numerical evaluation of Eq.~\eqref{eq:F_gau}, shown in Fig.~\ref{fig:FQ_vs_N}. The main plot clearly displays the superextensive behavior at the critical point.
This is the metrological counterpart of the monitored critical singularity and mirrors the nonanalytic QFI behavior found in other no-click transitions \cite{DiFresco2023MIPTMetrology,Zerba2023NoClick}.

\section{Finite-efficiency smoothing}
For any $q<1$, imperfect detection introduces a finite correlation length and destroys the singular monitored critical phase \cite{Paviglianiti2024Breakdown}. Close to the perfect-detection limit,
\begin{equation}\label{eq:xi_scaling}
\xi^{-1}=1-q+\mathcal{O}\big((1-q)^2\big),
\end{equation}
so the sharp momentum-space discontinuity of the perfect-detection steady state is replaced by a crossover of width set by $\xi^{-1}$. The detailed smoothing argument is given in Appendix~\ref{app:scaling_arguments}. The resulting singular $L^2$ contribution is cut off and replaced by the extensive law
\begin{equation}\label{eq:power}
F_{\mu\mu}\sim \frac{L}{1-q}.
\end{equation}
Thus, the monitored critical enhancement is suppressed by information loss: the ideal no-click singularity survives only as a diverging prefactor as $q\to 1^-$. This picture is fully consistent with the finite-length scale scenario established in Ref.~\cite{Paviglianiti2024Breakdown}. The scaling behavior in Eq.~\eqref{eq:power} is shown in the inset of Fig.~\ref{fig:FQ_vs_N}. The plot clearly shows that the power-law dependence of the QFI on $q$ also holds away from the limit $q \rightarrow 1$.


\section{Multiparameter interpretation of detector efficiency}
\begin{figure}
    \centering
    \includegraphics[width=0.9\linewidth]{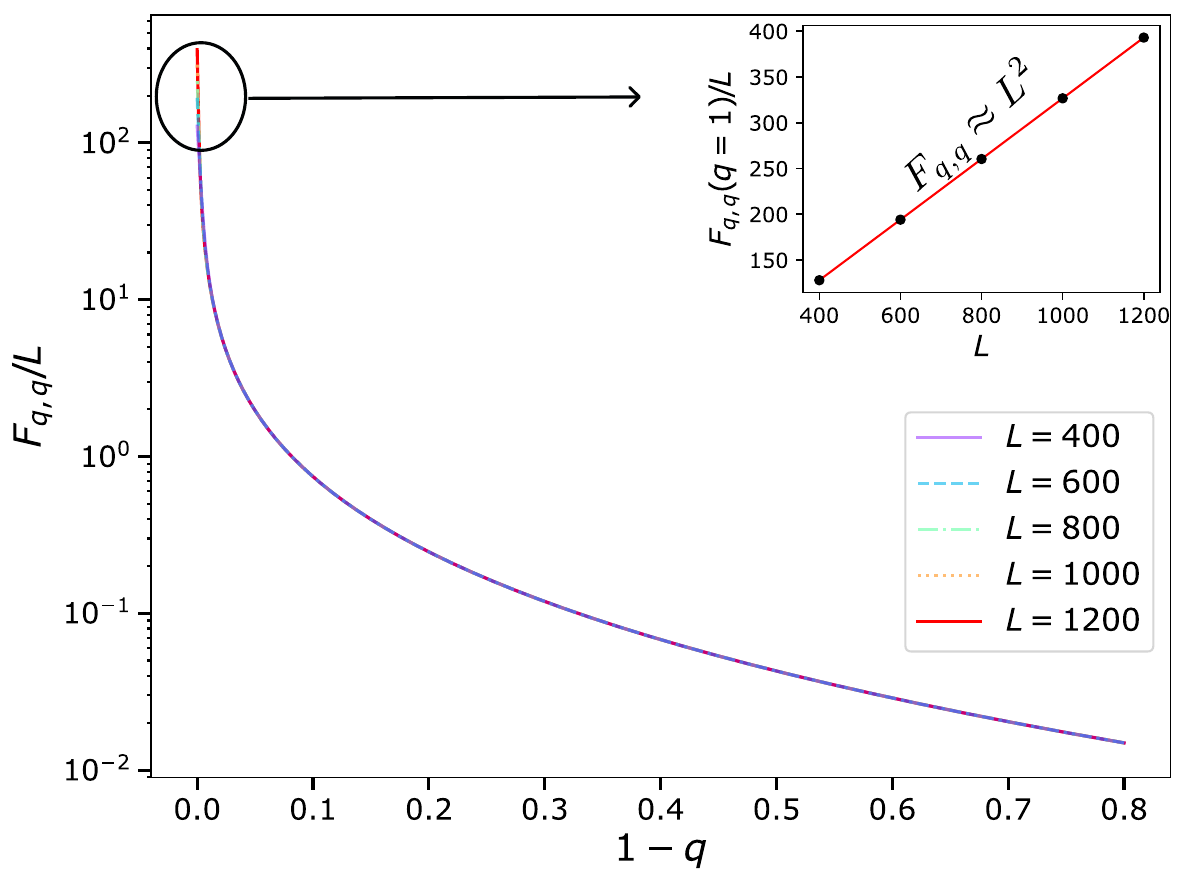}
    \caption{The main plot shows $F_{qq}$ as a function of $1-q$, clearly displaying the superextensive peak at $q=1$. The inset reports the scaling behavior of $F_{qq}/L$ at $q=1$, revealing its $L^2$ scaling. In these simulations, $\mu = 0$ and $\gamma = 0.3$.}
    \label{fig:FQ_vs_q}
\end{figure}

The detector efficiency can also be regarded as a sensing parameter in its own right. Its Fisher information can therefore be analyzed in the same way as $F_{\mu\mu}$. For any fixed $q<1$, the smoothing width remains finite and the corresponding Fisher information is extensive in system size. The singular limit $q\to 1^{-}$ is different: there the smoothing window collapses onto the momentum resolution scale, and the finite-size crossover restores the same $L^2$ enhancement found at perfect detection. The scaling argument is presented in Appendix~\ref{app:scaling_Fq}, while the corresponding numerical results are shown in Fig.~\ref{fig:FQ_vs_q}. In particular, the inset, which displays the scaling of $F_{qq}$ as $q\to 1^{-}$, clearly reveals the expected $L^2$ behavior.
\newline
In a quantum metrological setting involving more than one sensing parameter, such as $\mu$ and $q$, a natural question is whether the joint estimation problem is affected by quantum incompatibility. The relevant diagnostic is the mean Uhlmann curvature~\cite{Ragy2016Compatibility}. Using Eq.~\eqref{eq:MUC}, our numerical analysis reveals that the Uhlmann curvature $U_{\mu q}$ vanishes throughout the entire phase diagram. Consequently, the joint estimation problem for $\mu$ and $q$ remains compatible, implying that the quantum Cram\'er--Rao bound coincides with the ultimate multiparameter precision bound \cite{Ragy2016Compatibility,DiFresco2022Multiparameter}.
\newline
From a physical perspective, this compatibility is consistent with viewing the efficiency $q$ as an effective classical parameter, somewhat akin to temperature in equilibrium quantum systems. Indeed, a finite detection efficiency ($q<1$) makes it impossible to fully resolve individual trajectories, thereby introducing a probabilistic mixing of otherwise pure states that is reminiscent of thermal averaging. The analogy with finite temperature is, however, purely operational: whereas thermal mixedness originates from coupling to an equilibrium reservoir, the mixedness induced by imperfect detection arises from incomplete access to the quantum-jump record.
\newline
In this sense, imperfect detection acts as a compatible mixed-state parameter that suppresses the metrological enhancement associated with monitored criticality without introducing a genuinely quantum incompatibility bottleneck in the simultaneous estimation of $\mu$ and $q$. This interpretation is also consistent with the broader picture in which criticality can enhance sensitivity while reducing incompatibility \cite{DiFresco2022Multiparameter}.

\section{Conclusion}
We have investigated the quantum metrological properties of a continuously monitored Kitaev chain conditioned on an imperfect no-click record. Owing to the Gaussian nature of the conditional dynamics, the steady state is completely characterized by its fermionic correlation matrix, allowing an exact evaluation of both the quantum Fisher information and the mean Uhlmann curvature.
\newline
At perfect detection, the monitored steady state develops a singular structure at a critical momentum, leading to a super-extensive quantum Fisher information with respect to the chemical potential. For any $q<1$, incomplete detection rounds this singularity by introducing a finite smoothing length associated with the loss of trajectory information. Consequently, the Fisher information reverts to extensive scaling, with a prefactor that diverges as $q\rightarrow1^-$. At the same time, the vanishing of the mean Uhlmann curvature demonstrates that the joint estimation of $\mu$ and $q$ remains fully compatible throughout the phase diagram.
\newline
Although imperfect detection produces a mixed-state crossover reminiscent of thermal broadening, the underlying mechanism is fundamentally different. Finite temperature smooths equilibrium criticality through thermal occupation effects \cite{Zanardi2006MixedState,Zanardi2007BuresThermal}, whereas imperfect detection reflects incomplete information about the underlying quantum trajectories. Detector efficiency therefore constitutes a distinct information-loss parameter whose principal effect is to replace trajectory-resolved monitored criticality by a finite-length mixed-state crossover. These results establish a direct connection between monitored quantum metrology, information geometry, and measurement-induced critical phenomena \cite{Banchi2014Geometry,Carollo2017Uhlmann}.

\emph{Acknowledgement---}
GDF acknowledges support from the PNRR MUR project PE0000023--NQSTI. AC acknowledges support from European Union – Next Generation EU through projects: Eurostart 2022 Topological atom-photon interactions for quantum technologies MUR D.M. 737/2021 and PRIN 2022- PNRR no. P202253RLY Harnessing topological phases for quantum technologies. AC and DV acknowledge support from THENCE - Partenariato Esteso NQSTI - PE00000023 – Spoke 2.

\appendix

\section{Mode-resolved QFI for the $2\times 2$ correlation matrix}
\label{app:qfi_mode_decomposition}

Because Eq.~\eqref{eq:F_gau} is invariant under unitary transformations, the diagonal QFI component associated with the chemical potential can be evaluated independently for each momentum mode. One can then write
\begin{equation}\label{eq:Fisher_k}
F_{\mu\mu}=4\sum_k\sum_{ij}^{\prime}\frac{\left|\partial_{\mu}\Gamma_{ij}(k)\right|^2}{1-c_i(k)c_j(k)},
\end{equation}
where $c_i(k)$ are the eigenvalues of the $2\times 2$ correlation matrix $\Gamma(k)$.

For the present $2\times 2$ problem, it is convenient to write
\begin{equation}
\Gamma(k)=\vec{v}(k)\cdot\vec{\sigma},
\end{equation}
with $\vec{\sigma}$ the Pauli-vector basis and $c(k)=|\vec{v}(k)|$ the modulus of the eigenvalue of $\Gamma(k)$. The mode-resolved contribution then reads
\begin{equation}\label{eq:Fisher_vec}
F_{\mu\mu}(k)=4\left[\frac{2\big(\vec{v}\cdot\partial_\mu\vec{v}\big)^2}{\big(1-c^2\big)c^2}+\frac{2\left(|\partial_\mu\vec{v}|^2-\frac{\vec{v}\cdot\partial_\mu\vec{v}}{c^2}\right)}{1+c^2}\right].
\end{equation}
\section{Scaling arguments near perfect detection}
\label{app:scaling_arguments}

Two observations are important in the perfect-detection regime. First, for $q=1$, the first term in Eq.~\eqref{eq:Fisher_vec}, corresponding to the longitudinal component of $\partial_\mu\vec{v}$, vanishes because the state is perfectly pure. Second, as one approaches the critical mode $k^*$, the vector $\vec{v}$ exhibits a discontinuity of the form $\Delta\vec{v}=\left(2A^*a_0,0,0\right)$. Therefore, the behavior of $\vec{v}(k)$ in the vicinity of $k^*$ can be modeled as
\begin{equation}
\vec{v}(k)\approx \vec{v}_{-}(k)\Theta\big(k^*(\mu)-k\big)+\vec{v}_{+}(k)\Theta\big(k-k^*(\mu)\big),
\end{equation}
where $\Theta$ denotes the Heaviside step function. This implies that the critical behavior of the derivative is governed by
\begin{equation}
\partial_{\mu}\vec{v}(k)=-\frac{dk^*}{d\mu}\,\Delta\vec{v}\,\delta(k-k^*).
\end{equation}
For a finite-size system, the delta function can be regularized as $1/\Delta k$, with $\Delta k = 2\pi/L$ denoting the spacing of the momentum-space grid. Equivalently, the singular contribution is concentrated in a single momentum cell around $k^*$, so the number of relevant modes is of order one, while the weight of that cell scales as $1/\Delta k$. As a result, the expression for $F_{\mu\mu}$ for the critical mode takes the form
\begin{equation}
\frac{F_{\mu\mu}(k^*)}{4}=\left(\frac{dk^*}{d\mu}\right)^2\left(\frac{1}{\Delta k^2}\right)\left[\frac{2\left(|\Delta\vec{v}|^2-(\vec{v}_{+}\cdot\Delta\vec{v})\right)^2}{1+c^2}\right].
\end{equation}
This term dominates the expression in Eq.~\eqref{eq:Fisher_k} and leads to the scaling behavior quoted in Eq.~\eqref{eq:L2_scaling}.
\newline
For $q<1$, a finite correlation length is present in the system. Near the perfect-detection limit, the correlation length behaves as in Eq.~\eqref{eq:xi_scaling}, and $\vec{v}(k)$ can no longer be approximated by a sharp step function. Instead, it is convenient to write it as the sum of a smooth background contribution and a regularized step,
\begin{equation}\label{eq:v_q1}
\vec{v}(k)\approx \vec{v}_{\mathrm{bg}}(k)+\Delta\vec{v}\,f\left(\frac{k-k^*}{\delta k}\right),
\end{equation}
where $f$ denotes a smooth crossover function, and $\delta k$ is the momentum-space smoothing width around $k^*$, set by the inverse correlation length. For imperfect detection, $q<1$, this gives $\delta k \approx \xi^{-1} \approx 1-q$~\cite{Paviglianiti2024Breakdown}. The derivative then behaves as
\begin{equation}
\partial_{\mu}\vec{v}(k)\approx -\frac{dk^*}{d\mu}\,\Delta\vec{v}\,\frac{1}{\delta k}f'\left(\frac{k-k^*}{\delta k}\right).
\end{equation}
Hence the singular $1/\Delta k$ enhancement of the perfect-detection case is replaced by a smoothing-controlled factor $1/\delta k$, so that $|\partial_{\mu}\vec{v}|\propto (1-q)^{-1}$ close to $q=1$. The relevant number of contributing momentum modes is now the number of grid points inside the smoothing window around $k^*$, namely
\begin{equation}
N_{\mathrm{rel}}\sim \frac{\delta k}{\Delta k}.
\end{equation}
Since each contributing mode scales as $1/\delta k^2$, the full sum in Eq.~\eqref{eq:Fisher_k} behaves as
\begin{equation}
F_{\mu\mu}\sim N_{\mathrm{rel}}\frac{1}{\delta k^2}\sim \frac{1}{\Delta k\,\delta k}\sim \frac{L}{1-q},
\end{equation}
which yields the extensive scaling quoted in Eq.~\eqref{eq:power}.
\section{Scaling argument for $F_{qq}$}
\label{app:scaling_Fq}
The scaling argument for $F_{qq}$ follows the same reasoning as the one already used for $F_{\mu\mu}$. In particular, we start from Eq.~\eqref{eq:Fisher_vec} with the replacement $\mu\to q$. Near the smoothed critical region, it is convenient to use the same regularized form introduced in Appendix~\ref{app:scaling_arguments},
\begin{equation}
\vec{v}(k,q)\approx \vec{v}_{\mathrm{bg}}(k)+\Delta \vec{v}\,f\left(\frac{k-k^*}{\delta k(q)}\right),
\end{equation}
and to define the scaling variable
\begin{equation}
x=\frac{k-k^*}{\delta k(q)}.
\end{equation}
This allows us to write
\begin{equation}
\partial_q \vec{v}(k,q)\approx \Delta \vec{v}\, f'(x)\,\partial_q x.
\end{equation}
Since $k^*$ is independent of $q$,
\begin{equation}
\partial_q x=-\frac{k-k^*}{\delta k^2}\partial_q\delta k=-\frac{x}{\delta k}\partial_q\delta k.
\end{equation}
Close to the perfect-detection limit, $\delta k\approx 1-q$, so $\partial_q\delta k$ is of order one. Hence one obtains the scaling form
\begin{equation}
\partial_q \vec{v}(k,q)\sim \frac{1}{\delta k}\,\Phi\!\left(\frac{k-k^*}{\delta k}\right),
\end{equation}
where $\Phi$ is a localized function of order one. Therefore, each contributing mode gives
\begin{equation}
F_{qq}(k)\sim |\partial_q \vec{v}(k,q)|^2\sim \frac{1}{\delta k^2}\,\Psi\!\left(\frac{k-k^*}{\delta k}\right),
\end{equation}
with $\Psi$ another localized profile. As in Appendix~\ref{app:scaling_arguments}, the relevant number of contributing modes is
\begin{equation}
N_{\mathrm{rel}}\sim \frac{\delta k}{\Delta k}.
\end{equation}
For any fixed $q<1$, this yields the extensive behavior
\begin{equation}
F_{qq}\sim N_{\mathrm{rel}}\frac{1}{\delta k^2}\sim \frac{1}{\Delta k\,\delta k}\sim L.
\end{equation}
In the singular limit $q\to 1^{-}$, however, the smoothing width collapses until it becomes comparable with the momentum spacing, $\delta k\sim \Delta k\sim 1/L$. At that point only a finite number of critical modes contributes, and one recovers the superextensive scaling
\begin{equation}
F_{qq}\sim \frac{1}{\Delta k^2}\sim L^2.
\end{equation}
\section{Closed-form steady-state solution}
\label{app:closed_form_solution}

For completeness, we collect here the explicit algebraic solution of the steady-state Riccati equation~\eqref{eq:ARE} following Ref.~\cite{Paviglianiti2024Breakdown}. Writing the correlation matrix in the form of Eq.~\eqref{eq:Gamma_param}, we define
\begin{subequations}
\begin{align}
R(k)&=2\mu-2\cos k,\\
I(k)&=2\sin k,\\
S(k)&=\sqrt{\gamma^4+\big(R^2+I^2\big)^2+8\gamma^2\big[(1-4q+2q^2)I^2+R^2\big]}.
\end{align}
\end{subequations}
The two algebraic branches are then
\begin{subequations}
\begin{align}
a_{\pm}(k)&=\pm\frac{2\sqrt{2}R}{\sqrt{\gamma^2-4(R^2+I^2)+S}},\\
b_{\pm}(k)&=-\frac{1+a_{\pm}^2}{a_{\pm}}\frac{R}{I},\\
A_{\pm}(k)&=\frac{(1-q)\gamma a_{\pm}+2R}{2q\gamma(1+a_{\pm}^2)R}I.
\end{align}
\end{subequations}
The physical steady state corresponds to the branch that reproduces the proper limits as $q\to 0$ and $q\to 1$.

\bibliographystyle{apsrev4-2}
\bibliography{ref}

\end{document}